\documentclass[a4paper,fleqn,usenatbib,useAMS]{mnras}

\usepackage{graphicx}	
\usepackage{amsmath}	
\usepackage{amssymb}	
\usepackage{multicol}        
\usepackage{bm}		
\usepackage{pdflscape}	
\usepackage{natbib}
\usepackage{comment}
\usepackage{longtable}
\usepackage{array}
\usepackage[draft]{hyperref}

\title[L dwarf flares]{$K2$ Ultracool Dwarfs Survey. VI. White light superflares observed on an L5 dwarf and flare rates of L dwarfs}

\author[R. R. Paudel et al.]{R. R. Paudel$^{1,2,3}$\thanks{Contact e-mail: \href{mailto:rpaudel@udel.edu}{rpaudel@umbc.edu}},
 J. E. Gizis$^{3}$, 
 D. J. Mullan$^{3}$, 
 S. J. Schmidt$^{4}$, 
 A. J. Burgasser$^{5}$, 
\newauthor{P. K. G. Williams$^{6,7}$} 
\\
$^{1}$NASA Goddard Space Flight Center, Greenbelt, MD 20771, USA \\
$^{2}$University of Maryland, Baltimore County, 1000 Hilltop Cir, Baltimore, MD 21250, USA \\
$^{3}$Department of Physics and Astronomy, University of Delaware, Newark, DE, 19716, USA\\
$^{4}$Leibniz-Institute for Astrophysics Potsdam (AIP), An der Sternwarte 16, 14482, Potsdam, Germany \\
$^{5}$Center for Astrophysics and Space Science, University of California San Diego, La Jolla, CA 92093, USA\\
$^{6}$Harvard-Smithsonian Center for Astrophysics, 60 Garden Street, Cambridge, MA 02138, USA\\
$^{7}$American Astronomical Society, 1667 K Street NW Ste. 800, Washington, DC 20006, USA}

\pubyear{2018}

\begin{document}
\label{firstpage}
\pagerange{\pageref{firstpage}--\pageref{lastpage}}
\maketitle
\begin{abstract}
$Kepler$ $K2$ long cadence data are used to study white light flares in a sample of 45 L dwarfs. We identified 11 flares on 9 L dwarfs with equivalent durations of (1.3 - 198) hr and total (UV/optical/IR) energies of $\geq$0.9 $\times$ 10$^{32}$ erg. Two superflares with energies of $>$10$^{33}$ erg were detected on an L5 dwarf: this is the coolest object so far on which flares have been identified. The larger superflare on this L5 dwarf has an energy of 4.6$\times$ 10$^{34}$ ergs and an amplitude of $>$300 times the photospheric level: so far, this is the largest amplitude flare detected by the $Kepler/K2$ mission. The next coolest star on which we identified a flare was an L2 dwarf: 2MASS J08585891+1804463. Combining the energies of all the flares which we have identified on 9 L dwarfs with the total observation time which was dedicated by $Kepler$ to all 45 L dwarfs, we construct a composite flare frequency distribution (FFD). The FFD slope is quite shallow (-0.51$\pm$0.17), consistent with earlier results reported by Paudel et al. (2018) for one particular L0 dwarf, for which the FFD slope was found to be -0.34. Using the composite FFD, we predict that, in early and mid-L dwarfs, a superflare of energy 10$^{33}$ erg occurs every 2.4 years and a superflare of energy 10$^{34}$ erg occurs every 7.9 years. Analysis of our L dwarf flares suggests that magnetic fields of $\geq$0.13-1.3 kG are present on the stellar surface: such fields could suppress Type II radio bursts.
\end{abstract}
\newpage
\section{Introduction}
L dwarfs are ultracool dwarfs (hereafter UCDs) with effective temperatures $\lesssim$2300 K. Since the lowest mass hydrogen burning stars have spectral types as late as $\sim$L4, the early L dwarfs can be either young brown dwarfs or old hydrogen burning low mass stars. Objects with spectral types later than L4 are all brown dwarfs \citep{2000ApJ...542..464C,2008ApJ...689.1327S}. Empirically, L dwarfs have been found to have weak chromospheric emission lines, but some have strong radio-emissions. Furthermore, the coronae are usually not detected in these cool objects by currently available X-ray satellites. L dwarfs are rapidly rotating objects with periods $<$10 hours, but available data indicate that the L dwarfs do not follow the rotation-activity correlations (RAC) which are typical for stars with spectral types G, K, and M \citep{2008ApJ...684.1390R,2010ApJ...709..332B,2014ApJ...785....9W,2014ApJ...785...10C}. In this paper, we use $K2$ long cadence data to study the properties of white light flares (hereafter WLFs) of L dwarfs. In Section \ref{subsec:H_alpha}, \ref{subsec:X-ray and radio} and \ref{subsec:white light flares}, we give an overview of previous studies regarding the important magnetic activity indicators: H$\alpha$, X-ray and radio emission, and flares on L dwarfs. In Section \ref{subsection:flares on different stars}, we discuss about flares on stellar and substellar objects. 
\subsection{Previous studies: H$\alpha$ emission}
\label{subsec:H_alpha}
Extensive studies of H$\alpha$ emission from L dwarfs have already been reported by \cite{2015AJ....149..158S} and \cite{2016ApJ...826...73P}. \cite{2015AJ....149..158S} studied 551 L dwarfs using data obtained from SDSS, 2MASS and the WISE survey. Among the 551 targets, values of H$\alpha$ emission or upper limits were reported for 181 stars. The authors classified chromospherically active L dwarfs as those with H$\alpha$ equivalent width (EW) $>$0.75 \AA,  and chromospherically inactive L dwarfs as those with upper limits (non-detections) of H$\alpha$ EW $\leq$0.75 \AA. They found that $\sim$90\% of L0 dwarfs are chromospherically active, but this fraction decreases to $\sim$60\% in the case of L3 dwarfs. Likewise, the chromospheric activity fractions for the L4 and L5 dwarfs were found to be 33\% and 50\% respectively. The L6-L8 dwarfs have either very weak or no chromospheric activity. Likewise, there is a decline in the chromospheric activity (as measured by log $L_{\rm H\alpha}$/$L_{\rm bol}$) from a value of $\sim$ -3.8 in M0-M4 dwarfs to a value of -5.7 in the case of L3 dwarfs.\\ \\
\cite{2016ApJ...826...73P} studied H$\alpha$ emission from late-L and T dwarfs using optical (6300-9700 \AA) spectra obtained by the $Keck$ telescopes. In a sample of 109 L4-T8 dwarfs, they reported a detection rate of 9.2$\pm^{3.5}_{2.1}$\%. In particular, they detected H$\alpha$ emission in 9.3$\pm^{4.5}_{2.4}$\% of mid-to-late L dwarfs (L4-L8) in a sample of 75 objects. The sample of L dwarfs studied by \cite{2016ApJ...826...73P} might include binary systems. In view of that, it is possible that the overall detection rate of H$\alpha$ emission in single L dwarfs might be lower than they have reported. 
\subsection{Previous studies: X-ray and radio emission}
\label{subsec:X-ray and radio}
Although the L dwarfs are rapid rotators, their RACs  depart from those which have been obtained for stars earlier than M6.5 in the following sense. If the L dwarfs were to obey the RAC for GKM stars, the L dwarfs would be expected to have saturated  X-ray emission, with $L_{\rm X}$/$L_{\rm bol}$ $\approx$ 10$^{-3}$ \citep{2014ApJ...785...10C}. But this expectation is not observed. Instead, the L dwarfs show significantly lower levels of X-ray emission.\\ \\
Up to now, there have been only two detections of X-rays from L dwarfs. One detection refers to Kelu-1AB which is a binary system of two L dwarfs: Kelu-1A (L2$\pm$1) and Kelu-1B (L3.5$\pm$1) \citep{2005ApJ...634..616L}. \cite{2007A&A...471L..63A}  reported the detection of quiescent X-rays from Kelu-1AB with an X-ray luminosity of 2.9$^{+1.8}_{-1.3}$ $\times$ 10$^{25}$ erg s$^{-1}$ . The second detection refers to an L0-L1 star (J0331-27) which was found (in archival XMM data) to undergo a flare \citep{2020A&A...634L..13D}: at the peak of the flare, this star had an X-ray luminosity $L_{X}$ = 10$^{29.8}$ erg s$^{-1}$ in the energy range 0.5-2 keV. This luminosity is comparable to, or slightly larger than, the maximum luminosity observed in X-ray flares on M dwarfs with spectral types M6.5-M9.5. De Luca et al. searched for quiescent X-rays from J0331-27 in XMM and $Chandra$ data but found no significant detections of the source.\\ \\ 
Prior to the detection by De Luca et al., \cite{2014ApJ...785...10C} had compiled the largest catalog of L dwarfs with X-ray measurements: the catalog contains 10 L dwarfs with spectral types in the range L0-L8. Of these 10 stars, one had an X-ray detection, with log $L_{X}$/$L_{bol}$  = -4.7, while the remaining 9 had only upper limits of X-rays in the range log $L_{X}$/$L_{bol}$ $<$ -5.0 to -3.0. In addition to the L dwarfs reported by \cite{2014ApJ...785...10C}, we have independently searched for X-ray emission from another flaring L1 dwarf: WISEP J190648.47+401106.8 using $Chandra$ data. No X-ray photons were unambiguously detected from this L dwarf during an observation period of $\sim$50 ks (R. R. Paudel et al., in prep.).\\ \\
As regards the radio luminosity of the stars of interest to us, \cite{2010ApJ...709..332B} report that L dwarfs exhibit a pronounced deviation from the Guedel-Benz relation (GBR). The latter relation is based on observations of stars of spectral types F, G, K, and early-M: for such stars, $L_{X}$ has been found to be correlated linearly with the radio luminosity $L_{\nu,R}$ at frequency $\nu$: log $L_{\nu,R}$/$L_{X}$ $\sim$ -15.5 \citep{1993ApJ...405L..63G}. However, at a spectral type of M7-M9, the $L_{\nu,R}$/$L_{X}$ correlation is observed to break down “sharply” (\citealt{2010ApJ...709..332B}, especially their Fig. 8).  Though there is observed to be a decline in X-ray emission as we go to later L dwarfs, radio emission does not follow this decline. In some L dwarfs, $L_{\nu,R}$/$L_{X}$ is found to be several orders of magnitude larger than 10$^{-15.5}$ even when the stars are in the quiescent state (see also \citealt{2014ApJ...785....9W}: their Fig. 6). A possible explanation for the breakdown in GBR in L dwarfs has been proposed \citep{2010ApJ...721.1034M} based on the physics of magnetic reconnection. Two modes of reconnection are possible (slow and fast), depending on the process which determines the electrical conductivity in the ambient medium: if classical ohmic processes are dominant, reconnection occurs at a relatively slow rate. But in conditions where the Hall effect dominates, reconnection can occur orders of magnitude more rapidly. The latter case is believed to be the case on solar-like coronal conditions. But in L dwarfs, conditions may allow only slow reconnection to occur: in such cases, although X-rays will not be efficiently generated, the electrons emerging from the reconnection site can still have enough energy to be effective emitters of electron cyclotron maser radio emission. \\ \\
A compilation of the L dwarfs with radio emission can be found in \cite{2018haex.bookE.171W}. The number of L dwarfs with known radio emission is currently 9, among which 4 are confirmed to have radio emission that varies with periods of $\lesssim$ 1 hr time scales. Radio emission has been observed on objects with spectral type as late as T6.5 \citep{2016ApJ...818...24K}.
\subsection{Previous studies: Flares on L dwarfs}
\label{subsec:white light flares}
L dwarfs exhibit lower rates of optical flares than M dwarfs do (see for e.g. \citealt{2018ApJ...858...55P}). As a result it is very difficult to study the flare rates using ground-based telescopes with limited observation times. For example, \cite{2013MNRAS.428.2824K} and \cite{2015MNRAS.453.1484R} did not detect any flares on the L dwarfs which they monitored using ground-based telescopes.The precise photometry and continuous monitoring of the $K2$ mission make it relatively easy to study the WLF rates of L dwarfs. Using this opportunity, \cite{2013ApJ...779..172G} were able to detect 21 white light flares on the L1 dwarf WISEP J190648.47+401106.8 (hereafter W1906+40) by monitoring this object for three months in short cadence mode. The flares had (UV/visible/infrared) energies in the range $\sim$(10$^{29}$ - 10$^{32}$) erg.  W1906+40 was the first L dwarf on which WLFs were observed. Up to now, WLFs have been observed on 6 L dwarfs: W1906+40, SDSS J053341.43
+001434.1 (hereafter S0533+00; L0), SDSSp J005406.55-
003101.8 (L1), 2M1221+0257 (L0), 2M1232-0951 (L0) and ULAS J224940.13-011236.9 (hereafter U22-011; L2.5) \citep{2013ApJ...779..172G,2016ApJ...828L..22S,2017ApJ...838...22G,2018ApJ...858...55P,2019MNRAS.tmpL..42J}. All of these L dwarfs, except S0533+00 and U22-011, were monitored by the $Kepler$/$K2$ mission. Among all the WLFs observed on L dwarfs, the most powerful is the one observed on S0533+00 by the ground-based All Sky Automated Survey for SuperNovae (ASASSN). This flare had a total estimated bolometric energy of $>$6.2 $\times$ 10$^{34}$ erg \citep{2016ApJ...828L..22S}. Furthermore, U22-011 is the coolest flaring object detected so far: with spectral type L2.5, this means that it has $T_{\rm eff}$ $<$ 2000 K. It was observed by using the Next Generation Transit Survey (NGTS) and the flare had a total estimated energy of 3.4 $\times$ 10$^{33}$ erg \citep{2019MNRAS.tmpL..42J}.\\ \\ 
The noise level in the $K2$ long cadence data is $>$10\% for the fainter targets. As a result, small flares cannot be reliably detected. However, the flares with large amplitudes (relative to the photospheric level) can be easily detected even if the object is very faint and has very low photospheric emission. One such large flare with total (UV/visible/infrared) energy $\sim$10$^{33}$ erg was observed on the L1 dwarf SDSSp J005406.55-003101.8 by \cite{2017ApJ...838...22G} using long cadence data. 
\subsection{Flares on different classes of stars}
\label{subsection:flares on different stars}
It is well known that the Sun can produce flares in “active regions” where the local magnetic field is strong and stressed: the stressing comes about because of the motions of the gas in the photosphere due to convection. When the field stresses exceed a certain limit, the field relaxes to a lower energy state and some of the magnetic energy that was previously stored in the active region is converted rapidly into heat, bulk motion of gas, and energetic particles: this is a flare. The most common type of flare star has spectral class M: such stars have been observed to have strong magnetic fields \citep{2010MNRAS.407.2269M}. M stars also have extensive convective envelopes which can create stresses in the magnetic field. For this reason, it is believed that flares on M dwarfs also rely on the conversion of magnetic energy into heat, motion, and particles (e.g. \citealt{1989SoPh..121..239M}). Our interest in this paper is stars which are of spectral type L. The question is: does the internal structure of a star have any fundamental difference on the properties of the flares which are produced by different stars? \\ \\
To address this, we first consider the internal structure of stars of interest to us. The Sun has a radiative core with an outer envelope where convection transports energy. The energy owes its origin ultimately to proton burning in the core. Early-M dwarfs have internal structures such that proton burning occurs in a radiative core, and there is an extensive convective envelope: such a structure overlaps with the properties of the Sun. Stars with masses which are less than (roughly) 0.3-0.35 $M_{\odot}$ are completely convective on the main sequence (e.g. \citealt{2015ApJ...810L..18M}), but proton burning continues to occur in such stars.  However, for stars with masses $\lesssim$0.075 $M_{\odot}$ (corresponding to late-M on the main sequence), proton burning is no longer possible. They are also known as brown dwarfs: the only nuclear reactions in such objects (at least in their young stages) may involve deuterium burning. The early-L dwarfs are a mixture of both evolved old stars and young brown dwarfs. The internal structure of a star undoubtedly has an effect on the dynamo processes which can occur in that star, i.e. on the properties of the magnetic fields which can be generated (e.g. \citealt{2015ApJ...810L..18M}). In fact, observational properties of magnetic fields in M dwarfs show a range of properties \citep{2010MNRAS.407.2269M}: some stars have strong fields which are mainly dipolar, some have weak fields which are non-axisymmetric, and others have toroidal fields. As regards the strength of the fields on M dwarfs, the strongest surface fields are several kG \citep{2017ApJ...835L...4K,2017NatAs...1E.184S}: these upper limits on the field strength at the surface of a star may be constrained by equipartition with the photospheric gas pressure (\citealt{1996IAUS..176..237S}, especially his Fig. 3). Although there have been no reports so far about magnetic detections in L dwarfs, it would not be surprising to find that they also may contain a variety of magnetic topologies, but with field strengths which are also no larger than several kG. \\ \\
Turning now to a consideration of flares, although a star almost certainly generates its field in a medium where magnetic pressure is overwhelmed by gas pressure and/or gas kinetic energy, the processes of converting magnetic energy into non-magnetic forms almost certainly rely on plasma processes which occur in a medium where the magnetic field dominates the gas. Such a condition typically requires a medium of low density and high temperature, i.e. a stellar corona. In a region where magnetic flux tubes of opposite polarity can be brought into close proximity, reconnection of the field is an effective method of converting magnetic energy into the forms of energy which appear during a flare (e.g. \citealt{1964NASSP..50..425P}). The process of reconnection releases energy on a time-scale which is determined by the local physical conditions: specifically, the thickness of the current sheet which exists when a plasma is squeezed between two approaching magnetic flux tubes, and the electrical conductivity of the plasma. If ohmic  conductivity is at work, the rate of energy release is rather slow. However, in certain conditions, e.g. if Hall conductivity is relevant (i.e. ions decouple from the field lines but electrons do not), the rate of reconnection can increase by factors of order 10$^{6}$ \citep{2006ApJ...644L.145C}. The ratio between the thickness of a current sheet and certain basic length scales of the flare plasma helps to bring order to the properties of a large sample of flares in the Sun and in flare stars \citep{2008ApJ...676L..69C}.\\ \\
In view of the fact that reconnection is controlled by local properties in the corona, the flare process may be disconnected from the processes which generate the field within the star. To the extent that this is true, we expect that the properties of flares in stars of different internal structure need not be all that different. In this regard, \cite{2020A&A...634L..13D} have noted the following aspects of the X-ray flare which they discovered recently on an L dwarf: (i) the peak luminosity is similar to those observed on late-M dwarfs; (ii) the short decay time-scale of the flare, indicating that the flaring region on the star has a compact size, is consistent with flares on late-type M dwarfs;  (iii) the X-rays are emitted by a plasma which has a temperature of $\sim$16 MK, within the range of values reported  for late-M dwarf flares. De Luca et al. conclude: ``our observation shows that no qualitative change takes place in the properties of X-ray flares at the bottom of the main sequence down to $T_{\rm eff}$ $\sim$2100 K''. In fact, the flare temperature reported by De Luca et al. for the L-dwarf flare (with a total energy of 10$^{33.3}$ ergs in X-rays) is only slightly higher than the most likely temperature (12-13 MK) which occurs in a sample of 4500 smaller flares (with total energies of 10$^{29-31}$ ergs) on the Sun \citep{2011ApJ...736...75B}.\\ \\
These results suggest that flares in a range of stars from spectral type G to L share a number of properties, despite the difference in their internal structures.
\subsection{Plan of the present paper}
In this paper, we use photometry obtained by the $K2$ mission to study the light curves of 45 L dwarfs. Only the L dwarfs which can be distinguished from background in their target pixel level data are included in our sample. This is the largest sample of L dwarfs for which the flare rates have been studied. Our study will lead us to report for the first time the detection of flares on an L5 dwarf which produced two superflares during $K2$ Campaign 11. One of these flares is found to have the largest amplitude relative to the photospheric level and also the largest energy among all the L dwarf flares monitored by the $Kepler$/$K2$ mission.  
\subsection{Sample of L dwarfs observed by $K2$}
\begin{figure}
    \includegraphics[scale=0.35]{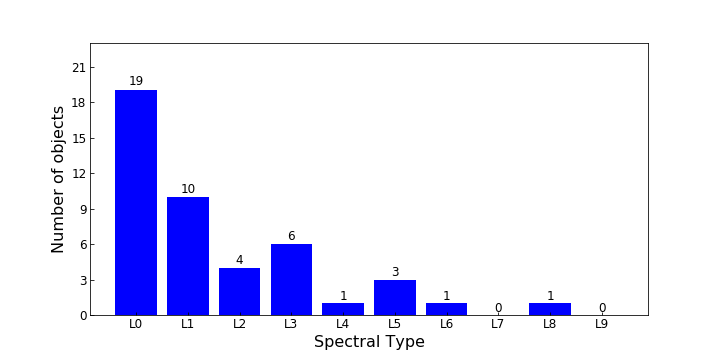} 
    \vspace{-5mm}
    \caption{Distribution of L dwarfs by spectral type.}
    \label{fig:histogram_Ldwarfs}
\end{figure}

We analyzed the light curves of 45 L dwarfs for which we obtained $K2$ photometry despite being very faint objects. In Table \ref{table:observation times}, we list the total observation times of L dwarfs in each $K2$ campaign. The total observation time of all L dwarfs in the sample is 4402.0 days and this is equal to 12.1 years.\\ \\ 
In Table \ref{table:list of L dwarfs}, we list the L dwarfs which were observed in various $K2$ campaigns with good photometry. The first column is the EPIC ID of each object and the second column is the 2MASS name of each object except for EPIC 236324763. Likewise, the third and fourth column give the information regarding the $K2$ campaign in which each object was observed and its optical spectral type respectively. One of the L dwarfs does not have an optical spectral type, so its near-infrared (NIR) spectral type is listed instead. The fifth column is the reference for the spectral type and the sixth column indicates whether the object was observed in short cadence mode or not \footnote{Here, we don't report any flares observed in short cadence mode only. We report only those which were observed in both long and short cadence. The results of short cadence data are either published in \cite{2018ApJ...858...55P} or will be published in Paudel et al. 2020, in prep.}. The total number of L dwarfs observed in short cadence mode is eleven. Figure \ref{fig:histogram_Ldwarfs} shows the distribution of L dwarfs in each spectral type.  
%
%
\begin{table} 
    \caption{\textbf{Total observation time by campaign}}
 	\label{table:observation times}
     \centering
     \begin{tabular}{ccc}
     \hline
      \textbf{Campaign \#} & \textbf{\# of L dwarfs} & \textbf{Time of Obs.} \\
       &  &  (d) \\
       \hline
       2 & 1 & 77.2\\
       3 & 0 & 0 \\
       4 & 2 & 144.5 \\
       5 & 16 & 1196.8 \\
       6 & 2 & 160.9 \\
       7 & 1 & 81.4 \\
       8 & 1 & 80.3 \\
       9 & 0 & 0 \\
       10 & 5 & 281.0 \\
       11 & 1 & 71.0  \\
       12 & 2 & 161 \\
       13 & 0 & 0 \\
       14 & 6 & 487.5 \\
       15 & 2 & 179.5 \\
       16 & 7 & 567.7 \\
       17 & 3 & 201.3 \\
       18 & 14 & 711.6 \\
       \hline
       \textbf{Total} & & 4402.0 \\
       \hline
    \end{tabular}
\end{table}
\begin{landscape}
\begin{table} 
    \caption{\textbf{Total observation time by campaign}}
 	\label{table:observation times}
     \centering
      \setlength\extrarowheight{-0.5pt}
     \begin{tabular}{cccccc}
     \hline
     	\textbf{EPIC} & \textbf{2MASS/other name} & \textbf{Cam. \#} & textbf{Sp. Type} & \textbf{ref} & \textbf{Short Cadence} \\ 
     	\hline 
     	204341806  & J16073799-2242468  & 2 & L0 & 1 \\
     	       210522262 & J04070752+1546457 & 4 & L3.5 & 2 \\
     	       210879793 & J04090950+2104393 &  4 & L3 & 3  \\
     	       211962038  &  J08264262+1939224  &  5, 18 &  L0  & 8 &  Yes \\
     	       211970944  &  J08175266+1947279   &  5, 18 & L0 & 2  & \\
     	       212102189  &  J08302724+2203456 &  5, 16, 18  & L0 & 2 \\
     	       211328277 & J08433323+1024470  &  5, 18  & L1 & 2 & Yes \\
     	       211357895 & J08503593+1057156 &  5, 18 & L6 & 4 \\
     	       211628806 & J0829066+145622  &  5, 18 & L2 & 3 & Yes \\
     	       211680042 & J08312221+1538511 & 5, 18 & L1 & 2 \\
     	       211727819 &  J08373282+1617380 & 5,18 & L0 & 2  \\
     	       211891128 & J08365239+1835455 & 5, 18 & L0 & 2 \\
     	       211963497 & J09094822+1940428 & 5 & L1 & 2 & Yes \\
     	       211978512 & J09053102+1954334 & 5 &  L0 & 2 \\
     	       212111554 & J08580549+2214582 & 5, 16, 18  & L1 & 2 \\
     	       212128548 &  J08564793+2235182 & 5,18 & L3 & 5  \\
     	       229227169 & J13530778-0857119 &  6 & L0 & 1 \\
     	       217976219 & J19090821-1937479 &  7 & L1 & 1 & Yes \\
     	       220186653 & J00540655-0031018 & 8 & L1 & 6 \\
     	       201181297 & 12130336-0432437  & 10 & L5 & 6 \\
     	       201299167 & J12025263-0227483 & 10  & L1  & 2 \\
     	       201482905 & J12035812+0015500 & 10 & L3 & 7  \\
     	       201658777  & J12212770+0257198  &  10 & L0 & 6  & Yes\\
     	       228730045 & J12321827-0951502 &  10 & L0 & 8  &  Yes  \\
     	       236324763 & VVV BD001 & 11 & L5 p & 12 \\
     	       246080803 & J2344062-073328  &  12 & L4.5 & 9  \\
     	       246303486 & J23255604-0259508 & 12 & L3 & 10  \\
     	       201528766  & J10501247+0058032  & 14 & L0  & 2  \\
     	       248442470  & J10484281+0111580  & 14 & L1 & 6 &  Yes\\
     	       248523311 & J10340564+0350164	& 14 & L0 & 2  \\
     	       248653486 &  J10431944+0712326 &  14 & L0  & 2 \\
     	       248862470  & J10433508+1213149  & 14 & L8 (NIR) & 10  \\
     	       248891072 & J10345117+1258407 &  14 & L0 & 2 \\
     	       249343675 & J15230657-2347526 &  15 & L0  & 14 & Yes \\
     	       249914869  & J1507476-162738 & 15 & L5  & 3 &  Yes \\
     	       211467731  & J08560211+1240150 & 16 & L0 &  2  \\
     	       211854467 & J08585891+1804463 & 16 & L2 & 2 \\
     	       212119590  &  J08354537+2224310 & 16 & L0 & 2  \\
     	       212127137  & J08535917+2233363 &  5,16, 18 & L2 & 2 \\
     	       251355936  & J0918382+213406 & 16 & L2.5 & 4 \\
     	       229227143  & J13530778-0857119  &  6, 17 & L0 &  1 \\
     	       251551345  & J13433872-0220446 & 17  & L1 & 2 \\
     	       251555071  &  J13334540-0215599 & 17 &  L3 & 2 &  Yes  \\
     	       211981633 & J08375977+1957279 & 5, 18 & L0 & 2 \\
     	       211602578  & J08403612+1434247 & 5, 18 & L1 & 2 \\
     	       \hline
     	\end{tabular} 
     	\\
     	 \textbf{References:}
     	       	1) \cite{2018ApJS..234....1B}; 2) \cite{2010AJ....139.1808S}; 3) \cite{2000AJ....120..447K} 
     	       	4) \cite{1999ApJ...519..802K}; 5) \cite{2003AJ....126.2421C}; 6) \cite{2015AJ....149..158S};  7) \cite{2007AJ....133.2258S}; 8) \cite{2014ApJ...794..143B};  9) \cite{2008ApJ...689.1295K};  10) \cite{2010ApJ...710.1142B}; 11) \cite{2014yCat.5144....0M}; 12) \cite{2013A&A...557L...8B}; 13) \cite{2012AJ....144..148G}; 14) \cite{2017MNRAS.465.4723K} 
\end{table}
\end{landscape}

\section{Data reduction and computations}
\subsection{$K2$ photometry}
We used the target pixel files (TPFs) of the objects and performed Point Spread Function (PSF) photometry to extract the corresponding light curves. The TPFs were obtained from the MAST archive. The PSF photometry is also known as Point Response Function (PRF). Under this method, a parameterized model developed by using the response of the pixels to stars of different brightness is fitted to the data. The PSF method is preferred over aperture photometry (AP) method for extracting the light curves of objects which lie in crowded regions of the sky \citep{lightkurve}. We used the PSF fitting model incorporated in the Python package `Lightkurve'
\citep{lightkurve}. The light curves were then detrended against any systematics (such as the instrumental noise) by using the K2 Systematics Correction (`$K2SC$', \citealt{2016MNRAS.459.2408A}).
\subsection{Flare detection}
We used the false discovery rate (FDR) analysis described in \cite{2001AJ....122.3492M} to identify possible flare candidates in the $K2SC$ detrended light curve. The same method was used by \cite{2012ApJ...754....4O} to identify flares in the light curves of their targets. For FDR analysis, we calculated the relative flux $F_{\rm rel,i}$ for each data point of the detrended light curve as follows:
\begin{equation}
F_{rel,i} = \frac{F_i - F_{m}}{F_{m}} 
\end{equation}
where \textit{F$_{i}$} is the flux in \textit{i}th epoch and $F_{\rm m}$ is the mean flux of the entire light curve of each target. Using the values of $F_{rel,i}$, we then calculated a statistic $\phi_{ij}$ for each consecutive observation epoch ($i,j$) as:
\begin{equation}
\phi_{ij} = \Big(\frac{F_{rel,i}}{\sigma_{i}} \Big) \times \Big(\frac{F_{rel,j}}{\sigma_{j}}\Big) , j = i+1
\end{equation}
Here \textit{$\sigma_{i}$} is the error in the flux which is associated with the \textit{i}th epoch. The possible flare candidates were first identified by estimating a critical threshold value of $\phi_{ij}$ as described in \cite{2001AJ....122.3492M}. The detailed information regarding this method of identifying flare candidates is given in \cite{2012ApJ...754....4O} and \cite{2018ApJ...858...55P}. The final flares were confirmed by inspecting each candidate by eye.
Among the sample of 45 L dwarfs, 9 were observed to flare. 
\subsection{Flare energy estimation}
To calculate the energies of our flares, we first measured the equivalent duration (ED) of each flare using the $K2$ light curve. ED has units of time and is the duration in which the flare produces the same amount of energy as the (sub)stellar object when it is in its quiescent state \citep{1972Ap&SS..19...75G}. We modeled each flare as a 10,000 K blackbody normalized to have the same count rate through the $K2$ response curve as the photosphere of the flaring L dwarf and estimated the energy of the flare emitted during an ED of 1s. This energy is estimated for UV/optical/IR wavelengths. We then multiplied this energy with the ED of each flare to get an estimate of the total energy emitted by the flare. A similar procedure for flare energy estimation is described in \cite{2017ApJ...838...22G,2017ApJ...845...33G}. The values of flare energy for an ED of 1s for each flaring L dwarf are given in Table \ref{table:ED1 of flaring L dwarfs}.
\begin{table} 
    \caption{\textbf{Flare energy for ED = 1 s for each flaring L dwarf}}
 	\label{table:ED1 of flaring L dwarfs}
     \centering
     \begin{tabular}{cc}
     \hline
       \textbf{EPIC} & \textbf{energy} \\
       & (erg) \\
       \hline
       211891128 & 4.6 $\times$ 10$^{28}$ \\
        220186653 & 5.3 $\times$ 10$^{28}$ \\
        201658777 & 7.4$\times$ 10$^{28}$ \\
        228730045 & 8.6$\times$ 10$^{28}$ \\
        211854467 & 5.8$\times$ 10$^{28}$ \\
       236324763 & 6.5$\times$ 10$^{28}$ \\
       201528766 & 2.9$\times$ 10$^{28}$ \\
       249343675 & 9.8$\times$ 10$^{28}$ \\
       212102189 & 8.7$\times$ 10$^{28}$ \\
      \hline
    \end{tabular}
\end{table}
\subsection{Flare on an L2 dwarf: 2MASS J08585891+1804463}
\begin{figure} 
   \centering
   \includegraphics[scale=0.6]{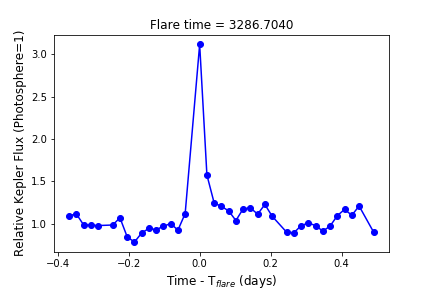}
    \caption{The flare observed on the L2 dwarf 2MASS J08585891+1804463. }
    \label{fig:L2 dwarf flare}
\end{figure}

2MASS J08585891+1804463 (hereafter 2M0858+1804) is an L2 dwarf \citep{ 2010AJ....139.1808S} and was observed by the $K2$ in Campaign 16 for 70.7 d. It has a photospheric (continuum) level of 71 counts s$^{-1}$ in the $K2$ light curve. A flare was observed on 2M0858+1804 at $Kepler$ time 3286.7039 during which the star brightened by a factor of $\sim$3  relative to the local photospheric level measured 10 hours before the flare. \footnote{There was a `thruster firing' of the spacecraft right before the peak flare time, but the centroid of the target remained in the same pixel with no significant change in the brightness level during that cadence.} The flare has an ED of 2.0 hr and lasted for 0.20 d. It has a total estimated energy of 4.2 $\times$ 10$^{32}$ erg. The flare light curve is shown in Figure \ref{fig:L2 dwarf flare}. The object 2M0858+1804 is the second L2 dwarf known to produce a WLF.
\subsection{Flares on an L5 dwarf: VVV BD001}
\begin{figure*}
    \includegraphics[width=15cm,height=6cm]{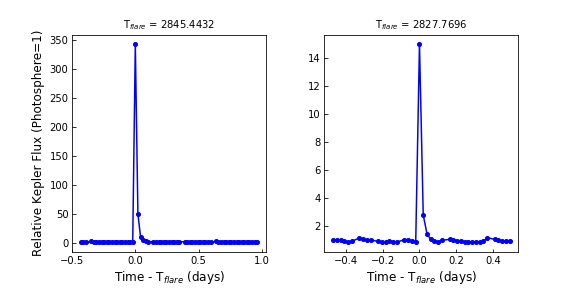}
    \caption{Flares observed on V01. The time along the $X$-axis is centered at peak flare time mentioned above each plot. The flux along the $Y$-axis is normalized by the median flux which corresponds to photospheric level.}
    \label{fig:flares of V01}
\end{figure*}

VVV BD001 (VVV J172640.2-273803, hereafter V01) is an L5 brown dwarf discovered by \cite{2013A&A...557L...8B} using the Vista Variables in the V\'ia L\'actea (VVV) survey \citep{2010NewA...15..433M,2012A&A...537A.107S}. It is an unusually blue L dwarf (UBLD) that is located at a distance of 18.5$\pm$0.14 pc, towards the Galactic bulge, and is a high proper motion object with $\mu_{\alpha}$  = -544.5$\pm$0.6 mas yr$^{-1}$ and $\mu_{\delta}$  = -326.4$\pm$0.4 mas yr$^{-1}$. This corresponds to a tangential velocity $V_{\rm tan}$ of 56.0$\pm$0.4 km s$^{-1}$. We note that, in a sample of 20 ``known blue L dwarfs'' which have unusually blue $J$ - $K_{s}$ colors, \cite{2010ApJS..190..100K} listed values of $V_{\rm tan}$ with a median value of 58$\pm$38 km s$^{-1}$. Thus, in terms of $V_{\rm tan}$, V01 does not appear at all exceptional when compared kinematically with the known blue L dwarfs. Previously, a sub-sample of 10 UBLD’s in the sample of Kirkpatrick et al. had been analyzed by \cite{2009AJ....137....1F} and had been found to have a median $V_{\rm tan}$ of 99$\pm$47 km s$^{-1}$: Kirkpatrick et al. noted that although the median $V_{\rm tan}$ = 58 km s$^{-1}$ of their expanded UBLD sample is “somewhat lower” than the median value reported by Faherty et al., still the result agrees with that of Faherty et al. within the 1$\sigma$ error bar. However, in a sample of 33 field L dwarfs, \cite{2004AJ....127.2948V} reported a median value of $V_{tan}$ = 24.5 km s$^{-1}$ : this indicates that V01 (with $V_{\rm tan}$ = 56 km s$^{-1}$) clearly has a larger tangential velocity than the field L dwarfs. It therefore seems likely that V01 belongs to an older population than the field L dwarfs.\\ \\
V01 is the first brown dwarf to be discovered in the very crowded region of sky where it is located. V01 was observed in long cadence mode by the $K2$ mission during Campaign 11 for $\sim$71 days. The median count rate of V01 in the $K2$ light curve is 122 counts s$^{-1}$. The photometric and physical properties of V01 are summarized in Table \ref{table:properties of V01}. \\ \\
Two superflares were detected by $K2$ on V01. The flare light curves are shown in Figure \ref{fig:flares of V01}. Following the detection of flares on the L2.5 dwarf U22-011 (in 2019), V01 is now the coolest object known to undergo flares. The larger of the two superflares occurred at $Kepler$ mission time 2845.4432 during which V01 brightened by more than 300 times above its photospheric level. The count rate at the peak of this flare is 41874 counts s$^{-1}$. This flare had an ED of 198 hr and lasted for 0.55 d. It is estimated that a 10,000 K flare with an ED of 1s has an (UV/optical/IR) energy of 6.5 $\times$ 10$^{28}$ erg. Using this estimate, the total (UV/optical/IR) energy emitted during the largest flare is estimated to be 4.6 $\times$ 10$^{34}$ erg.\\ \\
The weaker of the two superflares on V01 occurred at $Kepler$ mission time 2827.7696 during which V01 brightened by a factor of 15 times the photospheric level. The count rate at the peak of this flare is 1832 counts s$^{-1}$. It has an ED of 8.0 hr and lasted for 0.06 d. The total estimated (UV/optical/IR) energy of this flare is 2.0 $\times$ 10$^{33}$ erg.\\ \\
The fact that superflares have been detected on an unusually blue L5 star raises a question: is there anything unique about V01 compared to other stars in the UBLD class? \cite{2008ApJ...674..451B} suggested that UBLD’s differ from ``normal field L dwarfs'' in the sense that the condensate clouds in the photosphere of a UBLD may be either thinner, patchier, or contain larger grains than the clouds in a normal L dwarf of the same effective temperature. Burgasser et al. concluded that the UBLD’s may be ``relatively old'', and they suggested that the cloud properties might possibly be influenced by ``magnetic field strengths''. Magnetic fields are almost certainly at work in generating superflares on V01, but if fields are present, they might have other physical consequences as well, perhaps in determining the properties of the grains which are permitted to condense. As regards the term ``relatively old'' used by Burgasser et al. to describe UBLD’s, we note that \cite{2010ApJS..190..100K} also suggested, based on kinematics, that ``many of the blue L dwarfs may be old, though not as old as the halo population''. This still leaves a wide range of possible ages, since the inner halo of our galaxy has an age of 11.4 Gyr \citep{2012Natur.486...90K}. The fact that UBLD’s may have ages of several Gyr is not necessarily an argument against the possibility of magnetic activity: \cite{2008AJ....135..785W} have shown that the activity lifetimes of stars of spectral types M0-M7 increases towards later types, reaching as old as 8 Gyr (or so) at type M7.  If the trend of multi-Gyr activity lifetimes continues beyond M7, we expect that the L dwarfs could remain active even if they have ages of several Gyr.  In this regard, it is worth noting that the first L dwarf to be detected as the site of an X-ray flare (J0331-27) has IR colors which indicate ``that it is likely an older source'' with an age which may be $\geq$1 Gyr \citep{2020A&A...634L..13D}. Although J0331-27 was not present in the list of UBLD’s when Kirkpatrick et al. compiled their list in 2010, \cite{2020A&A...634L..13D} note that J0331-27 is ``bluer than field L1 dwarfs'': specifically, its $J$-$K_{s}$ color is 1.10$\pm$0.18, and this overlaps comfortably with the range of colors listed by \cite{2010ApJS..190..100K} for UBLD’s. \\ \\
Thus, we now know of two UBLD’s which have been detected as flaring objects. In this regard, we can answer the question posed at the start of the previous paragraph in the negative: V01 is $not$ necessarily unique among UBLD’s.\\ \\
%
%
%
%
The energies of the two flares in V01, i.e. 2.0 $\times$ 10$^{33}$ ergs and 4.6 $\times$ 10$^{34}$ ergs, are noteworthy in the context of the habitability of possible planets orbiting such stars. In this regard, we note that \cite{2018ApJ...867...71L} have discussed the photolysis effects of the UV radiation from a flare with total energy of 4 $\times$ 10$^{33}$ ergs, i.e. within the range of the energies of our two flares on V01. Loyd et al. conclude that the accumulated erosion of the atmosphere of an exoplanet in the habitable zone (HZ) of an M dwarf by flares with such energies ``could be significant over timescales of hundreds of megayears''. To be sure, the discussion of Loyd et al. applies to stars with somewhat earlier spectral types (early M), but their results may have applicability also to late L stars, where the HZ lies even closer to the parent star.  Moreover, another effect of large stellar flares is the generation of energetic proton fluxes, which may also lead to significant effects on the atmosphere of an exoplanet: \cite{2017ApJ...843...31Y} have suggested that energetic protons from M dwarf flares may lead to ``complete stripping of ozone from an Earth-like planet on timescales between 10$^{2}$ and 10$^{5}$ years''. Therefore, if the activity which we have discovered in V01 is typical of late-L dwarfs, the habitability of exoplanets orbiting late-L dwarfs may have to be called into question.  

\begin{table*}
 	\caption{\textbf{Properties of VVV BD001}}
 	\label{table:properties of V01}
 	\centering
     \begin{tabular*}{0.65\textwidth}{cccc}
     \hline
       & \textbf{Value} & \textbf{Units} & \textbf{Ref.} \\
       \hline
        \textbf{PHOTOMETRIC PROPERTIES} \\
       \hline
       Sp. Type & L5$\pm1$  &  & 1 \\
      \textit{J} & 13.27$\pm$0.02 & mag & 1 \\
      \textit{H} & 12.67$\pm$0.02 & mag & 1 \\
      \textit{K$_{s}$} & 12.20$\pm$0.02 & mag & 1 \\
      \textit{i}& 17.90$\pm$0.02 & mag & 2 \\
      \textit{G} & 18.23$\pm$0.00 & mag & 3 \\
      \textit{K$_{p}$} & 14.73 & mag & 4 \\
      \hline
      \textbf{PHYSICAL PARAMETERS}\\
       \hline
        $\alpha$ & 261.7$^{a}$ ($\pm$0.3 mas) & deg & 3 \\
        $\delta$ & -27.6$^{a}$ ($\pm$0.2 mas) & deg & 3 \\
        parallax & 54.0 $\pm$ 0.4 & mas & 3\\
        $\mu_{\alpha}$ & -544.5 $\pm$ 0.6 & mas yr$^{-1}$ & 3 \\
        $\mu_{\delta}$ & -326.4 $\pm$0.4 & mas yr$^{-1}$ & 3 \\
        \hline
        \end{tabular*}
        \\
        $^{a}$epoch J2015.5\\
        \textbf{References}: \\
(1) \cite{2013A&A...557L...8B}; (2) \cite{2016arXiv161205560C};\\
(3) \cite{2018arXiv180409365G}; (4) \cite{2017yCat.4034....0H}
\end{table*}
%
%
%
\section{Results}
We have identified a total of 11 flares on 45 L dwarfs observed in $K2$ long cadence mode. Most of the flares were identified on the early L dwarfs. Those flares have EDs in the range (1.3 - 198) hr and total estimated (UV/optical/IR) energies in the range (0.9 - 460) $\times$ 10$^{32}$ erg. More information regarding the ED, peak flare time, flare energy and flare duration of each flare is given in Table \ref{table:properties of flares of L dwarfs} \footnote{Flare energies of the targets which were observed in both short and long cadence modes are reported by using the EDs obtained from short cadence data to ensure more accuracy.}. In Table \ref{table:properties of flaring L dwarfs}, we list the 2MASS, PS1, $Gaia$ properties of the flaring L dwarfs. The EW of H$\alpha$ emission is also listed whenever available.\\ \\
We identified a strong flare on an L2 dwarf (2M0858+1804) and two superflares on an L5 dwarf (V01). So far, 2M0858+1804 is the second L2 dwarf on which a WLF has been observed. Likewise, V01 is the first L5 dwarf on which WLFs have been reported. Hence, V01 is currently the coolest substellar object that is known to produce WLFs. More interestingly, two superflares were identified on the L5 dwarf in a single $K2$ campaign in an interval of $\sim$18 days. The larger of these two superflares had an amplitude of $\sim$340 relative to the photospheric level and an ED of 198 hr. An amplitude of $\sim$340 times photospheric level is the largest amplitude among all the flares observed on the targets we have studied using $K2$ data.\\ \\
Assuming a similarity between the physical processes in solar flares and those in flares on (sub)stellar objects, a rough estimation of the maximum magnetic field strength $B$ associated with the flares can be obtained by using scaling relations between flare energy and magnetic field which have been reported by several authors:  \cite{2013A&A...549A..66A}, \cite{2013PASJ...65...49S}, and \cite{2019ApJ...876...58N}. We consider these three in turn.\\ \\
First, as regards flares on the Sun, \cite{2013A&A...549A..66A} make the following assumptions: the flare occurs when a certain fraction of the total available magnetic energy is released (e.g. as a result of reconnection) throughout an active region. The active region is assumed to have a characteristic linear dimension of $L_{bi}$ associated with the separation between bipolar spots of dominant positive and negative polarity. The magnetic field strength in each spot region on the solar surface is assumed to have a maximum value of $B_{m}$. In the simplest case, the portion of the field which is reconnected in order to power the flare is assumed to have a field strength which is a fraction $f'$ of $B_{m}$ throughout the flaring volume $V_{f}$ in the atmosphere. Also in the simplest case, the volume $V_{f}$ is assumed to equal some fraction $f''$ of the volume of a cube with sides of length $L_{bi}$. With these assumptions, the magnetic energy $M_{E}$ released per cm$^{3}$ is $M_{E}$ = $f'^{2}$ $B_{m}^{2}$/8$\pi$ ergs cm$^{-3}$ , while the magnetic energy released in the flare as a whole is $E_{f}$ = $M_{E}$ $\times$ $V_{f}$ = 0.04 $f'^{2}$ $f''$ $B_{m}^{2}$  $L_{bi}^{3}$ ergs. A particular numerical example is provided by Aulanier et al. They find that, by inserting values which may be (more or less) appropriate for conditions in the Sun, i.e. $B_{m}$ = 10$^{3}$ G and $L_{bi}$ = 5 $\times$ 10$^{9}$ cm, the flare energy is predicted (based on a model calculation which Aulanier et al. performed) to be 5 $\times$ 10$^{31}$ ergs. This is consistent with our expression for $E_{f}$ given above provided that the combination of terms 0.04 $f'^{2}$ $f''$ has the numerical value of 4 $\times$ 10$^{-4}$ , i.e. $f'^{2}$ $f''$ = 0.01. Provided that $f'$ and $f''$ do not differ by orders of magnitude, we can satisfy this relation by assuming that $f'$ and $f''$ are comparable, in which case $f'$ = $f''$ = 0.2. Thus, only about 20\% of the maximum available field strength needs to be reconnected, and only about 20\% of the available volume needs to participate in the flare event. In other words, we do not need to make extreme assumptions (e.g. by demanding that all of the magnetic energy throughout the entire volume of the active region be converted to flare energy) in order to account for the empirical properties of solar flares.\\ \\
Applying the scaling of \cite{2013A&A...549A..66A} to flare stars, we assume (in the absence of any firm evidence to the contrary), that the numerical values of the fractions $f'$ and $f''$ can be assumed to be essentially the same as in the Sun. In this context, the energies which are expected to be released in flares on a flare star can be written in terms of the star-spot magnetic field $B_{m}$ and the linear separation $L_{bi}$ between positive and negative spots on the stellar surface as follows: $E_{f}$ = 4 $\times$ 10$^{-4}$ $B_{m}^{2}$ $L_{bi}^{3}$. According to the results in Table \ref{table:properties of flares of L dwarfs} above, the observed values of $E_{f}$ are mainly in the range 10$^{32-34}$ ergs. In order to set a lower limit on $B_{m}$, we consider the limiting case in which the bipolar spots of an active region are placed as far apart as possible on the surface of a star: this upper limit on $L_{bi}$ would occur if the active region covered the entire surface of the star, with the positive spot at one pole of the star, and the negative spot at the opposite pole. It is physically impossible for the bipolar spots to be separated any farther than this. In this limiting case,  $L_{bi}$  $\approx$ $\pi$ $R_{\star}$ . With brown dwarf radii somewhat larger than Jupiter's radius, i.e. $R_{\star}$ $\approx$ 8 $\times$ $10^{9}$ cm, we find that, in the limiting case, we may set $L_{bi}$ $\approx$ 2.5 $\times$ 10$^{10}$ cm. Using this, in order to generate flare energies in the range 10$^{32-34}$ ergs (as reported in Table \ref{table:properties of flares of L dwarfs} of the present paper), we find the following lower limits on $B_{m}$ on our stars: $B_{m}$ $>$ 0.13$-$1.3 kiloGauss. In this case, because we have selected the two bipoles to be situated at both poles of the star, the field $B_{m}$ is truly to be considered as the global field of the star.\\ \\
Second, \cite{2013PASJ...65...49S} find that in a region of the Sun where the field strength is $B_{\rm k}$ (in units of kG), and where a fraction $f$ of the magnetic energy is released as flare energy, the amount of energy released in a flare (in units of ergs) amounts to $E_{f}$ $\approx$ 7 $\times$ 10$^{32}$ $B_{\rm k}^{2}$  ($f$/0.1) $L^{3}$. Here, the length $L$ is related to the area $A_{\rm spot}$ (in units of cm$^{2}$) of the sunspot in the active region where the flare occurs according to $L$ =  $\sqrt{A_{\rm spot}/10^{19}}$. The value of $10^{19}$ cm$^{2}$ which appears in this formula is the average area of a (large) spot on the Sun capable of generating the largest X10 flares ($\sim$10$^{32}$ erg): such a spot occupies an area which is of order 3 $\times$ 10$^{-4}$ times the area of the visible hemisphere of the Sun. Applying the formula for $E_{f}$ to a flare star, it is known that on such stars, spot sizes can occupy as much as 20\% of the area of the visible hemisphere (e.g. \citealt{1973MNRAS.164..343B}). Given that the radius of a flare star $R_{\star}$ is of order 8 $\times$ 10$^{9}$ cm, $A_{\rm spot}$ may be as large as 0.2 $\pi$ $R_{\star}^{2}$ $\approx$ 4 $\times$ 10$^{19}$ cm$^{2}$. This leads to $L$ = 2 in the above formula. Therefore, if $f$ retains a value of order 0.1 in flare stars, we find that the formula of Shibata et al. can be written as $E_{f}$ $\approx$ 5.6 $\times$ 10$^{33}$ $B_{\rm k}^{2}$.  In order to replicate the flare energies listed in Table \ref{table:properties of flares of L dwarfs} of the present paper, this formula suggests that we need to have field strengths $B_{\rm k}$ which lie in the range 0.13-1.3 kG. These values are similar to the estimates we have obtained from using the formula of \cite{2013A&A...549A..66A}. \\ \\
Third, \cite{2019ApJ...876...58N} suggest that the formula for $E_{f}$  obtained by \cite{2013PASJ...65...49S} should be regarded as an upper limit. To test this, they compare the Shibata et al. formula with the flares with maximum energy on each star among a sample of 279 stars which were detected in $Kepler$ as flaring objects. Visual inspection of their results (in their Figure 7) indicates that the upper limit on field strength in stars with large spot areal coverage ($>$1\% of the hemisphere) exceeds 3 kG in about 10\% of the stellar sample. However, when \cite{2019ApJ...876...58N} limit their sample of stars for which $Gaia$ reported reliable distances in DR2 \citep{2018A&A...616A..10G}, the number of solar-like stars in their sample decreased to 106: for these stars, only a few percent have upper limits as large as 3 kG. In most of the $Gaia$ stars, the mean value of the upper limit on surface field strength is of order 1 kG.\\ \\
Other data sets are also beginning to generate data which are relevant to the present discussion. E.g., $TESS$ data for Proxima Centauri \citep{2019ApJ...884..160V} indicate that with spot areal coverage of 5\% or more, the observed flare energies can be powered by fields in the range 1-2 kG.  And data from EVRYSCOPE \citep{2019arXiv190710735H} also indicate that flares on stars with spectral types ranging from K7 to M4 have energies such that, with spot areal coverages of at least 4-5\%, the most energetic flares can be powered by magnetic fields with strengths of 1.5 kG.\\ \\
In summary, fields of order 1 kG appear to be adequate to provide the magnetic energy which is needed to account for the upper limits of energies which are released in stellar flares.\\ \\
Is there any independent evidence that global fields of at least 0.13-1.3 kG are present on the surface of low mass flare stars? In this regard, we may cite magneto-convective models of low-mass flaring stars in which the magnetic field impedes the onset of convection: in these models, global structural changes occur in the star, including inflation of the stellar radius by several percent \citep{2001ApJ...559..353M}. In order to fit the empirical data for any particular star, an appropriate value must be assigned to the vertical magnetic field component which exists at the surface of the star. Specifically, to fit the data for the flare stars UV Ceti and its companion (with spectral types dM5.5 and dM6), it has been found that the global vertical field on these two stars must be 1.85 and 2.25 kG \citep{2018ApJ...860...15M}. And to fit the data for Trappist 1 (with spectral type M8), the models indicate that the global vertical field must have values in the range 1.45-1.70 kG \citep{2018ApJ...869..149M}. Thus, the interpretation we provided above of the observed flare energies in our sample of L dwarfs in terms of fields with strengths $B_{m}$ $\geq$ 0.13-1.3 kG is not inconsistent with the results of magneto-convective modeling of low-mass flaring stars with spectral types which reach almost as late as L0. \\ \\
\begin{table*}
 	 \centering
    \caption{\textbf{Properties of flares on L dwarfs}}
    \label{table:properties of flares of L dwarfs}
     \begin{tabular*}{0.65\textwidth}{cccccc}
     \hline
 \textbf{EPIC} & \textbf{Sp. Type} & \textbf{$T_{\rm peak}$} & \textbf{ED} & \textbf{Energy} & \textbf{duration} \\
       &  & (BJD - 2454833) & (hr) & (erg) & (d) \\
       \hline
       211891128 & L0 & 2377.0389 & 11.4 & 2.0 $\times$ 10$^{33}$ &  0.02 \\
        220186653 & L1 & 2595.7841 & 15.4 & 3.0 $\times$ 10$^{33}$ & 0.2 \\
        201658777 & L0 & 2799.9338 & 0.30 & 8.8 $\times$ 10$^{31}$ & 0.02 \\
        228730045 & L0 & 2811.7645 & 11.4 & 3.6 $\times$ 10$^{33}$ & 0.18 \\
        228730045 & L0 & 2755.0599 & 1.3 & 3.9 $\times$ 10$^{32}$ & 0.04 \\
        211854467 & L2 & 3286.7039 & 2.4 & 4.2 $\times$ 10$^{32}$ & 0.06 \\
       236324763 & L5 & 2845.4432  &  198 & 4.6 $\times$ 10$^{34}$ & 0.55 \\
       236324763 & L5 & 2827.7696 & 8.0 & 2.0 $\times$ 10$^{33}$ & 0.06 \\
       201528766 & L0 & 3138.0292 & 19.7 & 2.0 $\times$ 10$^{33}$ & 0.27 \\
       249343675 & L0 & 3238.3152 & 1.8 & 6.4 $\times$ 10$^{32}$ & 0.18 \\
       212102189 & L0 & 3427.1071 & 3.5 & 1.1 $\times$ 10$^{33}$ & 0.72 \\
      \hline
    \end{tabular*}
\end{table*}
\begin{table*}
    \centering
    \caption{\textbf{Properties of flaring L dwarfs}}
 	\label{table:properties of flaring L dwarfs}
     \begin{tabular*}{0.65\textwidth}{ccccccc}
     \hline
       \textbf{EPIC} & \textbf{\textit{J}} & \textbf{\textit{K$_{s}$}} & \textbf{\textit{i}} & \textbf{distance} & \textbf{H$\alpha$ EW} \\
       & (mag) & (mag) & (mag) & (pc) & (\AA) \\
       \hline
       211891128 & 16.51$\pm$0.12 & 15.18$\pm$0.10 & 20.8$\pm$0.1 & 67.3$\pm$13.0$^{a}$ \\
        220186653 & 15.73$\pm$0.05 & 14.38$\pm$0.07 & 20.07$\pm$0.04 & 51.9$\pm$3.3$^{b}$ & 5.0$^{e}$ \\
        201658777 & 13.17$\pm$0.02 & 11.95$\pm$0.03 & 17.44$\pm$0.01 &  18.5$^{c}$ & 6.1$^{f}$ \\
        228730045 & 13.73$\pm$0.03 & 12.55$\pm$0.03 & 18.04$\pm$0.00 & 26.4$\pm$4.9$^{d}$  \\
        211854467 & 16.35$\pm$0.10 & 15.14$\pm$0.12 & 20.80$\pm$0.03 & 76.4$\pm$14.7$^{a}$  &  \\
       236324763 & 13.27$\pm$0.02 & 12.20$\pm$0.02 & 17.90$\pm$0.02 & 18.5$\pm$0.1$^{b}$ \\
       201528766 & 16.58$\pm$0.14 & 15.78 & 20.39$\pm$0.02 & 54.0$\pm$10.4$^{a}$\\
       249343675 & 14.20$\pm$0.03 & 12.90$\pm$0.03 & 18.40$\pm$0.01 & 33.1$\pm$0.4$^{b}$   \\
       212102189 & 15.66$\pm$0.06 & 14.46$\pm$0.09 & 19.81$\pm$0.01 & 59.6$\pm$3.4$^{b}$ &  \\
      \hline
     \end{tabular*}
     \\
     \textbf{Notes:} \\
     i) $J$ and $K_{s}$ magnitudes are from 2MASS survey \citep{2003tmc..book.....C}. \\
     ii) $i$ magnitudes are from Pan-STARRS survey \citep{2016arXiv161205560C}.\\
     \textbf{References:}\\
     $^{a}$\cite{2010AJ....139.1808S};
     $^{b}$\cite{2018A&A...616A..10G};
     $^{c}$\cite{2018AJ....156...58B}; \\ $^{d}$\cite{2018ApJ...858...55P};
     $^{e}$\cite{2017ApJ...838...22G};
     $^{f}$\cite{2015AJ....149..158S}
\end{table*}
\subsection{Flare frequency distribution of L dwarfs}
\begin{figure} 
   \includegraphics[width=0.5\textwidth]{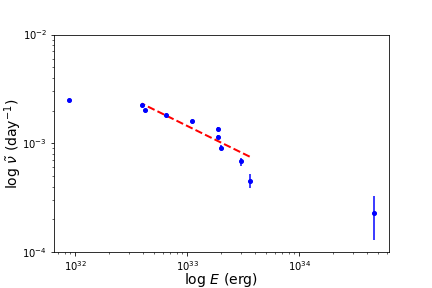}
   \vspace{-2mm}
    \caption{The FFD estimated by using the energies of all flares identified on L dwarfs observed by $K2$ long cadence mode. The kink seen around log $E$ $\sim$ 33.3 is not real. The errors were estimated using Poisson uncertainties.}
    \label{fig:FFD Ldwarfs}
\end{figure}

In Figure \ref{fig:FFD Ldwarfs}, we plot the flare frequency distribution (FFD) of L dwarfs using $K2$ long cadence data. This plot is useful in estimating the occurrence rate of flares with certain energies. Here, the energies of all flares observed on L0-L5 dwarfs are considered. In addition, the total time of observation (12.1 yrs) of all 45 L dwarfs is taken into account. The black dots represent the observed flare energies and the red dashed line is a weighted least-squares linear fit to power law distribution of flare energies. We used Poisson uncertainties to determine the weights for least square fitting. In addition, we did not include the minimum and maximum flare energies while fitting to reduce any bias due to those energies. The kink seen around log $E$ $\sim$ 33.3 erg is not real and is due to the closeness of the values of three flare energies around this value. The slope of the fitted line is -(0.51$\pm$0.17) and the cumulative frequency intercept at 10$^{30}$ erg is log $\tilde{\nu}_{30}$ = -1.42. Using these results, it is estimated that a superflare of energy 10$^{33}$ erg occurs every 2.4 years and a superflare of energy 10$^{34}$ erg occurs every 7.9 years on the L dwarfs.\\ \\
The slope of -0.51 of the FFD for L dwarfs is shallower compared to the slopes of FFD’s which have been observed in stars of earlier spectral types. For example, \cite{2018ApJ...858...55P}, in their study of  stars with spectral types M6.5 - M8.5 reported FFD slopes which were on average equal to –(0.58 - 0.70) regardless of stellar age.  However, in the case of one of their L0 dwarfs, they reported a shallower slope: -0.34. \cite{2018ApJ...854...14M} have argued that the shallowness of the slope for the FFD of an L0 star may be related to the low values of electrical conductivity which are expected in the atmosphere of such a cool star. This explanation could be helpful in understanding why the FFD for the L dwarfs reported in the present paper is also shallower than the averages reported for stars of earlier spectral types.

\section*{Acknowledgements}
The material in this paper is based upon work supported by NASA under award Nos. NNX15AV64G, NNX16AE55G and NNX16AJ22G. R.R.P acknowledges support from the University of Delaware Dissertation Fellowship program. Some/all of the data presented in this paper were obtained from the Mikulski Archive for Space Telescopes (MAST). STScI is operated by the Association of Universities for Research in Astronomy, Inc., under NASA contract NAS5-26555. This paper includes data collected by the Kepler mission. Funding for the Kepler mission is provided by the NASA Science Mission directorate. This work has made use of data from the European Space Agency (ESA) mission
{\it Gaia} (\url{https://www.cosmos.esa.int/gaia}), processed by the {\it Gaia}
Data Processing and Analysis Consortium (DPAC,
\url{https://www.cosmos.esa.int/web/gaia/dpac/consortium}). Funding for the DPAC
has been provided by national institutions, in particular the institutions
participating in the {\it Gaia} Multilateral Agreement. This research has made use of the VizieR catalogue access tool, CDS,
 Strasbourg, France. The original description of the VizieR service was
 published in A\&AS 143, 23. This work made use of the \url{http://gaia-kepler.fun}  crossmatch database created by Megan Bedell. \\

\textit{Software}: Python, IPython \cite{2007CSE.....9c..21P}, Astropy \citep{2013A&A...558A..33A}, Matplotlib \citep{Hunter:2007}, Numpy \citep{Oliphant:2015:GN:2886196}, Lightkurve \citep*{lightkurve}, K2SC \citep{2016MNRAS.459.2408A}, Jupyter \citep{Kluyver:2016aa}
%
%
\bibliographystyle{mnras}
\bibliography{astrobib}

\begin{thebibliography}{}
\makeatletter
\relax
\def\mn@urlcharsother{\let\do\@makeother \do\$\do\&\do\#\do\^\do\_\do\%\do\~}
\def\mn@doi{\begingroup\mn@urlcharsother \@ifnextchar [ {\mn@doi@}
  {\mn@doi@[]}}
\def\mn@doi@[#1]#2{\def\@tempa{#1}\ifx\@tempa\@empty \href
  {http://dx.doi.org/#2} {doi:#2}\else \href {http://dx.doi.org/#2} {#1}\fi
  \endgroup}
\def\mn@eprint#1#2{\mn@eprint@#1:#2::\@nil}
\def\mn@eprint@arXiv#1{\href {http://arxiv.org/abs/#1} {{\tt arXiv:#1}}}
\def\mn@eprint@dblp#1{\href {http://dblp.uni-trier.de/rec/bibtex/#1.xml}
  {dblp:#1}}
\def\mn@eprint@#1:#2:#3:#4\@nil{\def\@tempa {#1}\def\@tempb {#2}\def\@tempc
  {#3}\ifx \@tempc \@empty \let \@tempc \@tempb \let \@tempb \@tempa \fi \ifx
  \@tempb \@empty \def\@tempb {arXiv}\fi \@ifundefined
  {mn@eprint@\@tempb}{\@tempb:\@tempc}{\expandafter \expandafter \csname
  mn@eprint@\@tempb\endcsname \expandafter{\@tempc}}}

\bibitem[\protect\citeauthoryear{{Aigrain}, {Parviainen}  \& {Pope}}{{Aigrain}
  et~al.}{2016}]{2016MNRAS.459.2408A}
{Aigrain} S.,  {Parviainen} H.,   {Pope} B.~J.~S.,  2016, \mn@doi [\mnras]
  {10.1093/mnras/stw706}, \href
  {https://ui.adsabs.harvard.edu/abs/2016MNRAS.459.2408A} {459, 2408}

\bibitem[\protect\citeauthoryear{{Astropy Collaboration} et~al.,}{{Astropy
  Collaboration} et~al.}{2013}]{2013A&A...558A..33A}
{Astropy Collaboration} et~al., 2013, \mn@doi [\aap]
  {10.1051/0004-6361/201322068}, \href
  {http://adsabs.harvard.edu/abs/2013A%26A...558A..33A} {558, A33}

\bibitem[\protect\citeauthoryear{{Audard}, {Osten}, {Brown}, {Briggs},
  {G{\"u}del}, {Hodges-Kluck}  \& {Gizis}}{{Audard}
  et~al.}{2007}]{2007A&A...471L..63A}
{Audard} M.,  {Osten} R.~A.,  {Brown} A.,  {Briggs} K.~R.,  {G{\"u}del} M.,
  {Hodges-Kluck} E.,   {Gizis} J.~E.,  2007, \mn@doi [\aap]
  {10.1051/0004-6361:20078093}, \href
  {http://adsabs.harvard.edu/abs/2007A%26A...471L..63A} {471, L63}

\bibitem[\protect\citeauthoryear{{Aulanier}, {D{\'e}moulin}, {Schrijver},
  {Janvier}, {Pariat}  \& {Schmieder}}{{Aulanier}
  et~al.}{2013}]{2013A&A...549A..66A}
{Aulanier} G.,  {D{\'e}moulin} P.,  {Schrijver} C.~J.,  {Janvier} M.,  {Pariat}
  E.,   {Schmieder} B.,  2013, \mn@doi [\aap] {10.1051/0004-6361/201220406},
  \href {http://adsabs.harvard.edu/abs/2013A%26A...549A..66A} {549, A66}

\bibitem[\protect\citeauthoryear{{Bailer-Jones}, {Rybizki}, {Fouesneau},
  {Mantelet}  \& {Andrae}}{{Bailer-Jones} et~al.}{2018}]{2018AJ....156...58B}
{Bailer-Jones} C.~A.~L.,  {Rybizki} J.,  {Fouesneau} M.,  {Mantelet} G.,
  {Andrae} R.,  2018, \mn@doi [\aj] {10.3847/1538-3881/aacb21}, \href
  {http://adsabs.harvard.edu/abs/2018AJ....156...58B} {156, 58}

\bibitem[\protect\citeauthoryear{{Bardalez Gagliuffi} et~al.,}{{Bardalez
  Gagliuffi} et~al.}{2014}]{2014ApJ...794..143B}
{Bardalez Gagliuffi} D.~C.,  et~al., 2014, \mn@doi [\apj]
  {10.1088/0004-637X/794/2/143}, \href
  {http://adsabs.harvard.edu/abs/2014ApJ...794..143B} {794, 143}

\bibitem[\protect\citeauthoryear{{Baylor} et~al.,}{{Baylor}
  et~al.}{2011}]{2011ApJ...736...75B}
{Baylor} R.~N.,  et~al., 2011, \mn@doi [\apj] {10.1088/0004-637X/736/1/75},
  \href {https://ui.adsabs.harvard.edu/abs/2011ApJ...736...75B} {736, 75}

\bibitem[\protect\citeauthoryear{{Beam{\'{\i}}n} et~al.,}{{Beam{\'{\i}}n}
  et~al.}{2013}]{2013A&A...557L...8B}
{Beam{\'{\i}}n} J.~C.,  et~al., 2013, \mn@doi [\aap]
  {10.1051/0004-6361/201322190}, \href
  {http://adsabs.harvard.edu/abs/2013A%26A...557L...8B} {557, L8}

\bibitem[\protect\citeauthoryear{{Berger} et~al.,}{{Berger}
  et~al.}{2010}]{2010ApJ...709..332B}
{Berger} E.,  et~al., 2010, \mn@doi [\apj] {10.1088/0004-637X/709/1/332}, \href
  {http://adsabs.harvard.edu/abs/2010ApJ...709..332B} {709, 332}

\bibitem[\protect\citeauthoryear{{Best} et~al.,}{{Best}
  et~al.}{2018}]{2018ApJS..234....1B}
{Best} W.~M.~J.,  et~al., 2018, \mn@doi [\apjs] {10.3847/1538-4365/aa9982},
  \href {http://adsabs.harvard.edu/abs/2018ApJS..234....1B} {234, 1}

\bibitem[\protect\citeauthoryear{{Bopp} \& {Evans}}{{Bopp} \&
  {Evans}}{1973}]{1973MNRAS.164..343B}
{Bopp} B.~W.,  {Evans} D.~S.,  1973, \mn@doi [\mnras]
  {10.1093/mnras/164.4.343}, \href
  {https://ui.adsabs.harvard.edu/abs/1973MNRAS.164..343B} {164, 343}

\bibitem[\protect\citeauthoryear{{Burgasser}, {Looper}, {Kirkpatrick}, {Cruz}
  \& {Swift}}{{Burgasser} et~al.}{2008}]{2008ApJ...674..451B}
{Burgasser} A.~J.,  {Looper} D.~L.,  {Kirkpatrick} J.~D.,  {Cruz} K.~L.,
  {Swift} B.~J.,  2008, \mn@doi [\apj] {10.1086/524726}, \href
  {https://ui.adsabs.harvard.edu/abs/2008ApJ...674..451B} {674, 451}

\bibitem[\protect\citeauthoryear{{Burgasser}, {Cruz}, {Cushing}, {Gelino},
  {Looper}, {Faherty}, {Kirkpatrick}  \& {Reid}}{{Burgasser}
  et~al.}{2010}]{2010ApJ...710.1142B}
{Burgasser} A.~J.,  {Cruz} K.~L.,  {Cushing} M.,  {Gelino} C.~R.,  {Looper}
  D.~L.,  {Faherty} J.~K.,  {Kirkpatrick} J.~D.,   {Reid} I.~N.,  2010, \mn@doi
  [\apj] {10.1088/0004-637X/710/2/1142}, \href
  {http://adsabs.harvard.edu/abs/2010ApJ...710.1142B} {710, 1142}

\bibitem[\protect\citeauthoryear{{Cassak}, {Drake}  \& {Shay}}{{Cassak}
  et~al.}{2006}]{2006ApJ...644L.145C}
{Cassak} P.~A.,  {Drake} J.~F.,   {Shay} M.~A.,  2006, \mn@doi [\apjl]
  {10.1086/505690}, \href
  {https://ui.adsabs.harvard.edu/abs/2006ApJ...644L.145C} {644, L145}

\bibitem[\protect\citeauthoryear{{Cassak}, {Mullan}  \& {Shay}}{{Cassak}
  et~al.}{2008}]{2008ApJ...676L..69C}
{Cassak} P.~A.,  {Mullan} D.~J.,   {Shay} M.~A.,  2008, \mn@doi [\apjl]
  {10.1086/587055}, \href
  {https://ui.adsabs.harvard.edu/abs/2008ApJ...676L..69C} {676, L69}

\bibitem[\protect\citeauthoryear{{Chabrier}, {Baraffe}, {Allard}  \&
  {Hauschildt}}{{Chabrier} et~al.}{2000}]{2000ApJ...542..464C}
{Chabrier} G.,  {Baraffe} I.,  {Allard} F.,   {Hauschildt} P.,  2000, \mn@doi
  [\apj] {10.1086/309513}, \href
  {http://adsabs.harvard.edu/abs/2000ApJ...542..464C} {542, 464}

\bibitem[\protect\citeauthoryear{{Chambers} et~al.,}{{Chambers}
  et~al.}{2016}]{2016arXiv161205560C}
{Chambers} K.~C.,  et~al., 2016, preprint, \href
  {http://adsabs.harvard.edu/abs/2016arXiv161205560C} {} (\mn@eprint {arXiv}
  {1612.05560})

\bibitem[\protect\citeauthoryear{{Cook}, {Williams}  \& {Berger}}{{Cook}
  et~al.}{2014}]{2014ApJ...785...10C}
{Cook} B.~A.,  {Williams} P.~K.~G.,   {Berger} E.,  2014, \mn@doi [\apj]
  {10.1088/0004-637X/785/1/10}, \href
  {http://adsabs.harvard.edu/abs/2014ApJ...785...10C} {785, 10}

\bibitem[\protect\citeauthoryear{{Cruz}, {Reid}, {Liebert}, {Kirkpatrick}  \&
  {Lowrance}}{{Cruz} et~al.}{2003}]{2003AJ....126.2421C}
{Cruz} K.~L.,  {Reid} I.~N.,  {Liebert} J.,  {Kirkpatrick} J.~D.,   {Lowrance}
  P.~J.,  2003, \mn@doi [\aj] {10.1086/378607}, \href
  {http://adsabs.harvard.edu/abs/2003AJ....126.2421C} {126, 2421}

\bibitem[\protect\citeauthoryear{{Cutri} et~al.,}{{Cutri}
  et~al.}{2003}]{2003tmc..book.....C}
{Cutri} R.~M.,  et~al., 2003, {2MASS All Sky Catalog of point sources.}

\bibitem[\protect\citeauthoryear{{De Luca} et~al.,}{{De Luca}
  et~al.}{2020}]{2020A&A...634L..13D}
{De Luca} A.,  et~al., 2020, \mn@doi [\aap] {10.1051/0004-6361/201937163},
  \href {https://ui.adsabs.harvard.edu/abs/2020A&A...634L..13D} {634, L13}

\bibitem[\protect\citeauthoryear{{Faherty}, {Burgasser}, {Cruz}, {Shara},
  {Walter}  \& {Gelino}}{{Faherty} et~al.}{2009}]{2009AJ....137....1F}
{Faherty} J.~K.,  {Burgasser} A.~J.,  {Cruz} K.~L.,  {Shara} M.~M.,  {Walter}
  F.~M.,   {Gelino} C.~R.,  2009, \mn@doi [\aj] {10.1088/0004-6256/137/1/1},
  \href {https://ui.adsabs.harvard.edu/abs/2009AJ....137....1F} {137, 1}

\bibitem[\protect\citeauthoryear{{Gaia Collaboration}, {Brown}, {Vallenari},
  {Prusti}, {de Bruijne}, {Babusiaux}  \& {Bailer-Jones}}{{Gaia Collaboration}
  et~al.}{2018a}]{2018arXiv180409365G}
{Gaia Collaboration} {Brown} A.~G.~A.,  {Vallenari} A.,  {Prusti} T.,  {de
  Bruijne} J.~H.~J.,  {Babusiaux} C.,   {Bailer-Jones} C.~A.~L.,  2018a,
  preprint, \href {http://adsabs.harvard.edu/abs/2018arXiv180409365G} {}
  (\mn@eprint {arXiv} {1804.09365})

\bibitem[\protect\citeauthoryear{{Gaia Collaboration} et~al.,}{{Gaia
  Collaboration} et~al.}{2018b}]{2018A&A...616A..10G}
{Gaia Collaboration} et~al., 2018b, \mn@doi [\aap]
  {10.1051/0004-6361/201832843}, \href
  {http://adsabs.harvard.edu/abs/2018A%26A...616A..10G} {616, A10}

\bibitem[\protect\citeauthoryear{{Gershberg}}{{Gershberg}}{1972}]{1972Ap&SS..19...75G}
{Gershberg} R.~E.,  1972, \mn@doi [\apss] {10.1007/BF00643168}, \href
  {http://adsabs.harvard.edu/abs/1972Ap%26SS..19...75G} {19, 75}

\bibitem[\protect\citeauthoryear{{Gizis}, {Burgasser}, {Berger}, {Williams},
  {Vrba}, {Cruz}  \& {Metchev}}{{Gizis} et~al.}{2013}]{2013ApJ...779..172G}
{Gizis} J.~E.,  {Burgasser} A.~J.,  {Berger} E.,  {Williams} P.~K.~G.,  {Vrba}
  F.~J.,  {Cruz} K.~L.,   {Metchev} S.,  2013, \mn@doi [\apj]
  {10.1088/0004-637X/779/2/172}, \href
  {http://adsabs.harvard.edu/abs/2013ApJ...779..172G} {779, 172}

\bibitem[\protect\citeauthoryear{{Gizis}, {Paudel}, {Schmidt}, {Williams}  \&
  {Burgasser}}{{Gizis} et~al.}{2017a}]{2017ApJ...838...22G}
{Gizis} J.~E.,  {Paudel} R.~R.,  {Schmidt} S.~J.,  {Williams} P.~K.~G.,
  {Burgasser} A.~J.,  2017a, \mn@doi [\apj] {10.3847/1538-4357/aa6197}, \href
  {http://adsabs.harvard.edu/abs/2017ApJ...838...22G} {838, 22}

\bibitem[\protect\citeauthoryear{{Gizis}, {Paudel}, {Mullan}, {Schmidt},
  {Burgasser}  \& {Williams}}{{Gizis} et~al.}{2017b}]{2017ApJ...845...33G}
{Gizis} J.~E.,  {Paudel} R.~R.,  {Mullan} D.,  {Schmidt} S.~J.,  {Burgasser}
  A.~J.,   {Williams} P.~K.~G.,  2017b, \mn@doi [\apj]
  {10.3847/1538-4357/aa7da0}, \href
  {http://adsabs.harvard.edu/abs/2017ApJ...845...33G} {845, 33}

\bibitem[\protect\citeauthoryear{{Griffith} et~al.,}{{Griffith}
  et~al.}{2012}]{2012AJ....144..148G}
{Griffith} R.~L.,  et~al., 2012, \mn@doi [\aj] {10.1088/0004-6256/144/5/148},
  \href {http://adsabs.harvard.edu/abs/2012AJ....144..148G} {144, 148}

\bibitem[\protect\citeauthoryear{{Guedel} \& {Benz}}{{Guedel} \&
  {Benz}}{1993}]{1993ApJ...405L..63G}
{Guedel} M.,  {Benz} A.~O.,  1993, \mn@doi [\apjl] {10.1086/186766}, \href
  {https://ui.adsabs.harvard.edu/abs/1993ApJ...405L..63G} {405, L63}

\bibitem[\protect\citeauthoryear{{Howard}, {Corbett}, {Law}, {Ratzloff},
  {Glazier}, {Fors}, {del Ser}  \& {Haislip}}{{Howard}
  et~al.}{2019}]{2019arXiv190710735H}
{Howard} W.~S.,  {Corbett} H.,  {Law} N.~M.,  {Ratzloff} J.~K.,  {Glazier} A.,
  {Fors} O.,  {del Ser} D.,   {Haislip} J.,  2019, arXiv e-prints, \href
  {https://ui.adsabs.harvard.edu/abs/2019arXiv190710735H} {p. arXiv:1907.10735}

\bibitem[\protect\citeauthoryear{{Huber}, {Bryson}  \& {et al.}}{{Huber}
  et~al.}{2017}]{2017yCat.4034....0H}
{Huber} D.,  {Bryson} S.~T.,   {et al.} 2017, VizieR Online Data Catalog, \href
  {http://adsabs.harvard.edu/abs/2017yCat.4034....0H} {4034}

\bibitem[\protect\citeauthoryear{Hunter}{Hunter}{2007}]{Hunter:2007}
Hunter J.~D.,  2007, \mn@doi [Computing In Science \& Engineering]
  {10.1109/MCSE.2007.55}, 9, 90

\bibitem[\protect\citeauthoryear{{Jackman} et~al.,}{{Jackman}
  et~al.}{2019}]{2019MNRAS.tmpL..42J}
{Jackman} J.~A.~G.,  et~al., 2019, \mn@doi [\mnras] {10.1093/mnrasl/slz039},
  \href {http://adsabs.harvard.edu/abs/2019MNRAS.tmpL..42J} {}

\bibitem[\protect\citeauthoryear{{Kalirai}}{{Kalirai}}{2012}]{2012Natur.486...90K}
{Kalirai} J.~S.,  2012, \mn@doi [\nat] {10.1038/nature11062}, \href
  {https://ui.adsabs.harvard.edu/abs/2012Natur.486...90K} {486, 90}

\bibitem[\protect\citeauthoryear{{Kao}, {Hallinan}, {Pineda}, {Escala},
  {Burgasser}, {Bourke}  \& {Stevenson}}{{Kao}
  et~al.}{2016}]{2016ApJ...818...24K}
{Kao} M.~M.,  {Hallinan} G.,  {Pineda} J.~S.,  {Escala} I.,  {Burgasser} A.,
  {Bourke} S.,   {Stevenson} D.,  2016, \mn@doi [\apj]
  {10.3847/0004-637X/818/1/24}, \href
  {http://adsabs.harvard.edu/abs/2016ApJ...818...24K} {818, 24}

\bibitem[\protect\citeauthoryear{{Kirkpatrick} et~al.,}{{Kirkpatrick}
  et~al.}{1999}]{1999ApJ...519..802K}
{Kirkpatrick} J.~D.,  et~al., 1999, \mn@doi [\apj] {10.1086/307414}, \href
  {http://adsabs.harvard.edu/abs/1999ApJ...519..802K} {519, 802}

\bibitem[\protect\citeauthoryear{{Kirkpatrick} et~al.,}{{Kirkpatrick}
  et~al.}{2000}]{2000AJ....120..447K}
{Kirkpatrick} J.~D.,  et~al., 2000, \mn@doi [\aj] {10.1086/301427}, \href
  {http://adsabs.harvard.edu/abs/2000AJ....120..447K} {120, 447}

\bibitem[\protect\citeauthoryear{{Kirkpatrick} et~al.,}{{Kirkpatrick}
  et~al.}{2008}]{2008ApJ...689.1295K}
{Kirkpatrick} J.~D.,  et~al., 2008, \mn@doi [\apj] {10.1086/592768}, \href
  {http://adsabs.harvard.edu/abs/2008ApJ...689.1295K} {689, 1295}

\bibitem[\protect\citeauthoryear{{Kirkpatrick} et~al.,}{{Kirkpatrick}
  et~al.}{2010}]{2010ApJS..190..100K}
{Kirkpatrick} J.~D.,  et~al., 2010, \mn@doi [\apjs]
  {10.1088/0067-0049/190/1/100}, \href
  {https://ui.adsabs.harvard.edu/abs/2010ApJS..190..100K} {190, 100}

\bibitem[\protect\citeauthoryear{Kluyver et~al.,}{Kluyver
  et~al.}{2016}]{Kluyver:2016aa}
Kluyver T.,  et~al., 2016, in Loizides F.,  Schmidt B.,  eds, Positioning and
  Power in Academic Publishing: Players, Agents and Agendas. pp 87 -- 90

\bibitem[\protect\citeauthoryear{{Kochukhov} \& {Lavail}}{{Kochukhov} \&
  {Lavail}}{2017}]{2017ApJ...835L...4K}
{Kochukhov} O.,  {Lavail} A.,  2017, \mn@doi [\apjl]
  {10.3847/2041-8213/835/1/L4}, \href
  {https://ui.adsabs.harvard.edu/abs/2017ApJ...835L...4K} {835, L4}

\bibitem[\protect\citeauthoryear{{Koen}}{{Koen}}{2013}]{2013MNRAS.428.2824K}
{Koen} C.,  2013, \mn@doi [\mnras] {10.1093/mnras/sts208}, \href
  {http://adsabs.harvard.edu/abs/2013MNRAS.428.2824K} {428, 2824}

\bibitem[\protect\citeauthoryear{{Koen}, {Miszalski}, {V{\"a}is{\"a}nen}  \&
  {Koen}}{{Koen} et~al.}{2017}]{2017MNRAS.465.4723K}
{Koen} C.,  {Miszalski} B.,  {V{\"a}is{\"a}nen} P.,   {Koen} T.,  2017, \mn@doi
  [\mnras] {10.1093/mnras/stw3106}, \href
  {https://ui.adsabs.harvard.edu/\#abs/2017MNRAS.465.4723K} {465, 4723}

\bibitem[\protect\citeauthoryear{{Liu} \& {Leggett}}{{Liu} \&
  {Leggett}}{2005}]{2005ApJ...634..616L}
{Liu} M.~C.,  {Leggett} S.~K.,  2005, \mn@doi [\apj] {10.1086/496915}, \href
  {http://adsabs.harvard.edu/abs/2005ApJ...634..616L} {634, 616}

\bibitem[\protect\citeauthoryear{{Loyd} et~al.,}{{Loyd}
  et~al.}{2018}]{2018ApJ...867...71L}
{Loyd} R.~O.~P.,  et~al., 2018, \mn@doi [\apj] {10.3847/1538-4357/aae2bd},
  \href {https://ui.adsabs.harvard.edu/abs/2018ApJ...867...71L} {867, 71}

\bibitem[\protect\citeauthoryear{{MacDonald}, {Mullan}  \&
  {Dieterich}}{{MacDonald} et~al.}{2018}]{2018ApJ...860...15M}
{MacDonald} J.,  {Mullan} D.~J.,   {Dieterich} S.,  2018, \mn@doi [\apj]
  {10.3847/1538-4357/aac2c0}, \href
  {https://ui.adsabs.harvard.edu/abs/2018ApJ...860...15M} {860, 15}

\bibitem[\protect\citeauthoryear{{Mace}}{{Mace}}{2014}]{2014yCat.5144....0M}
{Mace} G.~N.,  2014, VizieR Online Data Catalog, \href
  {http://adsabs.harvard.edu/abs/2014yCat.5144....0M} {5144}

\bibitem[\protect\citeauthoryear{{Miller} et~al.,}{{Miller}
  et~al.}{2001}]{2001AJ....122.3492M}
{Miller} C.~J.,  et~al., 2001, \mn@doi [\aj] {10.1086/324109}, \href
  {http://adsabs.harvard.edu/abs/2001AJ....122.3492M} {122, 3492}

\bibitem[\protect\citeauthoryear{{Minniti} et~al.,}{{Minniti}
  et~al.}{2010}]{2010NewA...15..433M}
{Minniti} D.,  et~al., 2010, \mn@doi [\na] {10.1016/j.newast.2009.12.002},
  \href {http://adsabs.harvard.edu/abs/2010NewA...15..433M} {15, 433}

\bibitem[\protect\citeauthoryear{{Morin}, {Donati}, {Petit}, {Delfosse},
  {Forveille}  \& {Jardine}}{{Morin} et~al.}{2010}]{2010MNRAS.407.2269M}
{Morin} J.,  {Donati} J.~F.,  {Petit} P.,  {Delfosse} X.,  {Forveille} T.,
  {Jardine} M.~M.,  2010, \mn@doi [\mnras] {10.1111/j.1365-2966.2010.17101.x},
  \href {https://ui.adsabs.harvard.edu/abs/2010MNRAS.407.2269M} {407, 2269}

\bibitem[\protect\citeauthoryear{{Mullan}}{{Mullan}}{1989}]{1989SoPh..121..239M}
{Mullan} D.~J.,  1989, \mn@doi [\solphys] {10.1007/BF00161698}, \href
  {https://ui.adsabs.harvard.edu/abs/1989SoPh..121..239M} {121, 239}

\bibitem[\protect\citeauthoryear{{Mullan}}{{Mullan}}{2010}]{2010ApJ...721.1034M}
{Mullan} D.~J.,  2010, \mn@doi [\apj] {10.1088/0004-637X/721/2/1034}, \href
  {http://adsabs.harvard.edu/abs/2010ApJ...721.1034M} {721, 1034}

\bibitem[\protect\citeauthoryear{{Mullan} \& {MacDonald}}{{Mullan} \&
  {MacDonald}}{2001}]{2001ApJ...559..353M}
{Mullan} D.~J.,  {MacDonald} J.,  2001, \mn@doi [\apj] {10.1086/322336}, \href
  {https://ui.adsabs.harvard.edu/abs/2001ApJ...559..353M} {559, 353}

\bibitem[\protect\citeauthoryear{{Mullan} \& {Paudel}}{{Mullan} \&
  {Paudel}}{2018}]{2018ApJ...854...14M}
{Mullan} D.~J.,  {Paudel} R.~R.,  2018, \mn@doi [\apj]
  {10.3847/1538-4357/aaa960}, \href
  {http://adsabs.harvard.edu/abs/2018ApJ...854...14M} {854, 14}

\bibitem[\protect\citeauthoryear{{Mullan}, {Houdebine}  \&
  {MacDonald}}{{Mullan} et~al.}{2015}]{2015ApJ...810L..18M}
{Mullan} D.~J.,  {Houdebine} E.~R.,   {MacDonald} J.,  2015, \mn@doi [\apjl]
  {10.1088/2041-8205/810/2/L18}, \href
  {https://ui.adsabs.harvard.edu/abs/2015ApJ...810L..18M} {810, L18}

\bibitem[\protect\citeauthoryear{{Mullan}, {MacDonald}, {Dieterich}  \&
  {Fausey}}{{Mullan} et~al.}{2018}]{2018ApJ...869..149M}
{Mullan} D.~J.,  {MacDonald} J.,  {Dieterich} S.,   {Fausey} H.,  2018, \mn@doi
  [\apj] {10.3847/1538-4357/aaee7c}, \href
  {https://ui.adsabs.harvard.edu/abs/2018ApJ...869..149M} {869, 149}

\bibitem[\protect\citeauthoryear{{Notsu} et~al.,}{{Notsu}
  et~al.}{2019}]{2019ApJ...876...58N}
{Notsu} Y.,  et~al., 2019, \mn@doi [\apj] {10.3847/1538-4357/ab14e6}, \href
  {https://ui.adsabs.harvard.edu/abs/2019ApJ...876...58N} {876, 58}

\bibitem[\protect\citeauthoryear{Oliphant}{Oliphant}{2015}]{Oliphant:2015:GN:2886196}
Oliphant T.~E.,  2015, Guide to NumPy, 2nd edn.
CreateSpace Independent Publishing Platform, USA

\bibitem[\protect\citeauthoryear{{Osten}, {Kowalski}, {Sahu}  \&
  {Hawley}}{{Osten} et~al.}{2012}]{2012ApJ...754....4O}
{Osten} R.~A.,  {Kowalski} A.,  {Sahu} K.,   {Hawley} S.~L.,  2012, \mn@doi
  [\apj] {10.1088/0004-637X/754/1/4}, \href
  {http://adsabs.harvard.edu/abs/2012ApJ...754....4O} {754, 4}

\bibitem[\protect\citeauthoryear{{Paudel}, {Gizis}, {Mullan}, {Schmidt},
  {Burgasser}, {Williams}  \& {Berger}}{{Paudel}
  et~al.}{2018}]{2018ApJ...858...55P}
{Paudel} R.~R.,  {Gizis} J.~E.,  {Mullan} D.~J.,  {Schmidt} S.~J.,  {Burgasser}
  A.~J.,  {Williams} P.~K.~G.,   {Berger} E.,  2018, \mn@doi [\apj]
  {10.3847/1538-4357/aab8fe}, \href
  {http://adsabs.harvard.edu/abs/2018ApJ...858...55P} {858, 55}

\bibitem[\protect\citeauthoryear{{Perez} \& {Granger}}{{Perez} \&
  {Granger}}{2007}]{2007CSE.....9c..21P}
{Perez} F.,  {Granger} B.~E.,  2007, \mn@doi [Computing in Science and
  Engineering] {10.1109/MCSE.2007.53}, \href
  {http://adsabs.harvard.edu/abs/2007CSE.....9c..21P} {9, 21}

\bibitem[\protect\citeauthoryear{{Petschek}}{{Petschek}}{1964}]{1964NASSP..50..425P}
{Petschek} H.~E.,  1964, {Magnetic Field Annihilation}.
p.~425

\bibitem[\protect\citeauthoryear{{Pineda}, {Hallinan}, {Kirkpatrick}, {Cotter},
  {Kao}  \& {Mooley}}{{Pineda} et~al.}{2016}]{2016ApJ...826...73P}
{Pineda} J.~S.,  {Hallinan} G.,  {Kirkpatrick} J.~D.,  {Cotter} G.,  {Kao}
  M.~M.,   {Mooley} K.,  2016, \mn@doi [\apj] {10.3847/0004-637X/826/1/73},
  \href {http://adsabs.harvard.edu/abs/2016ApJ...826...73P} {826, 73}

\bibitem[\protect\citeauthoryear{{Ramsay}, {Hakala}  \& {Doyle}}{{Ramsay}
  et~al.}{2015}]{2015MNRAS.453.1484R}
{Ramsay} G.,  {Hakala} P.,   {Doyle} J.~G.,  2015, \mn@doi [\mnras]
  {10.1093/mnras/stv1742}, \href
  {http://adsabs.harvard.edu/abs/2015MNRAS.453.1484R} {453, 1484}

\bibitem[\protect\citeauthoryear{{Reiners} \& {Basri}}{{Reiners} \&
  {Basri}}{2008}]{2008ApJ...684.1390R}
{Reiners} A.,  {Basri} G.,  2008, \mn@doi [\apj] {10.1086/590073}, \href
  {http://adsabs.harvard.edu/abs/2008ApJ...684.1390R} {684, 1390}

\bibitem[\protect\citeauthoryear{{Saar}}{{Saar}}{1996}]{1996IAUS..176..237S}
{Saar} S.~H.,  1996, in {Strassmeier} K.~G.,  {Linsky} J.~L.,  eds,  IAU
  Symposium Vol. 176, Stellar Surface Structure. p.~237

\bibitem[\protect\citeauthoryear{{Saito} et~al.,}{{Saito}
  et~al.}{2012}]{2012A&A...537A.107S}
{Saito} R.~K.,  et~al., 2012, \mn@doi [\aap] {10.1051/0004-6361/201118407},
  \href {http://adsabs.harvard.edu/abs/2012A%26A...537A.107S} {537, A107}

\bibitem[\protect\citeauthoryear{{Saumon} \& {Marley}}{{Saumon} \&
  {Marley}}{2008}]{2008ApJ...689.1327S}
{Saumon} D.,  {Marley} M.~S.,  2008, \mn@doi [\apj] {10.1086/592734}, \href
  {http://adsabs.harvard.edu/abs/2008ApJ...689.1327S} {689, 1327}

\bibitem[\protect\citeauthoryear{{Schmidt}, {Cruz}, {Bongiorno}, {Liebert}  \&
  {Reid}}{{Schmidt} et~al.}{2007}]{2007AJ....133.2258S}
{Schmidt} S.~J.,  {Cruz} K.~L.,  {Bongiorno} B.~J.,  {Liebert} J.,   {Reid}
  I.~N.,  2007, \mn@doi [\aj] {10.1086/512158}, \href
  {http://adsabs.harvard.edu/abs/2007AJ....133.2258S} {133, 2258}

\bibitem[\protect\citeauthoryear{{Schmidt}, {West}, {Hawley}  \&
  {Pineda}}{{Schmidt} et~al.}{2010}]{2010AJ....139.1808S}
{Schmidt} S.~J.,  {West} A.~A.,  {Hawley} S.~L.,   {Pineda} J.~S.,  2010,
  \mn@doi [\aj] {10.1088/0004-6256/139/5/1808}, \href
  {http://adsabs.harvard.edu/abs/2010AJ....139.1808S} {139, 1808}

\bibitem[\protect\citeauthoryear{{Schmidt}, {Hawley}, {West}, {Bochanski},
  {Davenport}, {Ge}  \& {Schneider}}{{Schmidt}
  et~al.}{2015}]{2015AJ....149..158S}
{Schmidt} S.~J.,  {Hawley} S.~L.,  {West} A.~A.,  {Bochanski} J.~J.,
  {Davenport} J.~R.~A.,  {Ge} J.,   {Schneider} D.~P.,  2015, \mn@doi [\aj]
  {10.1088/0004-6256/149/5/158}, \href
  {http://adsabs.harvard.edu/abs/2015AJ....149..158S} {149, 158}

\bibitem[\protect\citeauthoryear{{Schmidt} et~al.,}{{Schmidt}
  et~al.}{2016}]{2016ApJ...828L..22S}
{Schmidt} S.~J.,  et~al., 2016, \mn@doi [\apjl] {10.3847/2041-8205/828/2/L22},
  \href {http://adsabs.harvard.edu/abs/2016ApJ...828L..22S} {828, L22}

\bibitem[\protect\citeauthoryear{{Shibata} et~al.,}{{Shibata}
  et~al.}{2013}]{2013PASJ...65...49S}
{Shibata} K.,  et~al., 2013, \mn@doi [\pasj] {10.1093/pasj/65.3.49}, \href
  {https://ui.adsabs.harvard.edu/abs/2013PASJ...65...49S} {65, 49}

\bibitem[\protect\citeauthoryear{{Shulyak}, {Reiners}, {Engeln}, {Malo},
  {Yadav}, {Morin}  \& {Kochukhov}}{{Shulyak}
  et~al.}{2017}]{2017NatAs...1E.184S}
{Shulyak} D.,  {Reiners} A.,  {Engeln} A.,  {Malo} L.,  {Yadav} R.,  {Morin}
  J.,   {Kochukhov} O.,  2017, \mn@doi [Nature Astronomy]
  {10.1038/s41550-017-0184}, \href
  {http://adsabs.harvard.edu/abs/2017NatAs...1E.184S} {1, 0184}

\bibitem[\protect\citeauthoryear{{Vida}, {Ol{\'a}h}, {K{\H{o}}v{\'a}ri}, {van
  Driel-Gesztelyi}, {Mo{\'o}r}  \& {P{\'a}l}}{{Vida}
  et~al.}{2019}]{2019ApJ...884..160V}
{Vida} K.,  {Ol{\'a}h} K.,  {K{\H{o}}v{\'a}ri} Z.,  {van Driel-Gesztelyi} L.,
  {Mo{\'o}r} A.,   {P{\'a}l} A.,  2019, \mn@doi [\apj]
  {10.3847/1538-4357/ab41f5}, \href
  {https://ui.adsabs.harvard.edu/abs/2019ApJ...884..160V} {884, 160}

\bibitem[\protect\citeauthoryear{Vin{\'\i}cius, Barentsen, Hedges,
  Gully-Santiago  \& Cody}{Vin{\'\i}cius et~al.}{2018}]{lightkurve}
Vin{\'\i}cius Z.,  Barentsen G.,  Hedges C.,  Gully-Santiago M.,   Cody A.~M.,
  2018, KeplerGO/lightkurve, \mn@doi{10.5281/zenodo.1181928}, \url
  {http://doi.org/10.5281/zenodo.1181928}

\bibitem[\protect\citeauthoryear{{Vrba} et~al.,}{{Vrba}
  et~al.}{2004}]{2004AJ....127.2948V}
{Vrba} F.~J.,  et~al., 2004, \mn@doi [\aj] {10.1086/383554}, \href
  {https://ui.adsabs.harvard.edu/abs/2004AJ....127.2948V} {127, 2948}

\bibitem[\protect\citeauthoryear{{West}, {Hawley}, {Bochanski}, {Covey},
  {Reid}, {Dhital}, {Hilton}  \& {Masuda}}{{West}
  et~al.}{2008}]{2008AJ....135..785W}
{West} A.~A.,  {Hawley} S.~L.,  {Bochanski} J.~J.,  {Covey} K.~R.,  {Reid}
  I.~N.,  {Dhital} S.,  {Hilton} E.~J.,   {Masuda} M.,  2008, \mn@doi [\aj]
  {10.1088/0004-6256/135/3/785}, \href
  {https://ui.adsabs.harvard.edu/\#abs/2008AJ....135..785W} {135, 785}

\bibitem[\protect\citeauthoryear{{Williams}}{{Williams}}{2018}]{2018haex.bookE.171W}
{Williams} P.~K.~G.,  2018, {Radio Emission from Ultracool Dwarfs}.
p.~171, \mn@doi{10.1007/978-3-319-55333-7_171}

\bibitem[\protect\citeauthoryear{{Williams}, {Cook}  \& {Berger}}{{Williams}
  et~al.}{2014}]{2014ApJ...785....9W}
{Williams} P.~K.~G.,  {Cook} B.~A.,   {Berger} E.,  2014, \mn@doi [\apj]
  {10.1088/0004-637X/785/1/9}, \href
  {http://adsabs.harvard.edu/abs/2014ApJ...785....9W} {785, 9}

\bibitem[\protect\citeauthoryear{{Youngblood} et~al.,}{{Youngblood}
  et~al.}{2017}]{2017ApJ...843...31Y}
{Youngblood} A.,  et~al., 2017, \mn@doi [\apj] {10.3847/1538-4357/aa76dd},
  \href {http://adsabs.harvard.edu/abs/2017ApJ...843...31Y} {843, 31}

\makeatother
\end{thebibliography}


\bsp	
\label{lastpage}
\end{document}